\documentclass[
aps,%
12pt,%
final,%
notitlepage,%
oneside,%
onecolumn,%
nobibnotes,%
nofootinbib,%
superscriptaddress,%
noshowpacs,%
centertags]%
{revtex4}%
\usepackage{geometry}
\usepackage{graphics}

\begin{document}
\selectlanguage{english}

\title{Charge asymmetry in the photonic 
production of charmed mesons}
\author{\firstname{A.~V.}~\surname{Berezhnoy}}
\email{aber@ttk.ru}
\affiliation{SINP MSU, Moscow, Russia}%
\author{\firstname{A.~K.}~\surname{Likhoded}}
\email{likhoded@mx.ihep.su}
\affiliation{IHEP, Protvino, Russia}

\begin{abstract}
Charge asymmetries for the charm meson production 
($D^{*+}$--$D^{*-}$,  $D^{*0}$--$\bar D^{*0}$
 and  
$D^+_s$--$\bar D^-_s$) have been estimated for the COMPASS kinematic
conditions in the framework of perturbative recombination model. 
Mass corrections have been taken into account in the calculations. 
The large  asymmetry for $D^+_s$--$\bar D^-_s$ production has been predicted.
\end{abstract}

\maketitle

\section{Introduction}

A production  of hadrons is a good testbed 
for the research of quark and gluon interactions
at the distance of color confinement.

On the one hand the intermediate value of $c$-quark mass 
complicates the process description in the  QCD frame work, 
on the other hand
this quark mass value allows us to research heavy quark 
interaction at the distances, 
which are larger than there  Compton wave length. 

The production asymmetry of charmed particles and antiparticles
has been observed in many high-energy experiment.
 For the first time such asymmetry had been discovered in  the
charm hadronic production~\cite{hadronic_asym}.	The asymmetry  value
in that process depends on the quark structure of initial hadrons, and
can be naturally explained by taking into account the interaction with the
 hadronic remnant~\cite{remnant_model}. 
 
 It is more amazing that a such charge asymmetry exists 
in the charm photoproduction in the photon fragmentation 
region~\cite{E691, NA14, E687}.
E691~\cite{E691} and 
E687 \cite{E687} Collaborations observed 
the statistically significant asymmetry value
for  $D^+ - D^-$, $D^0 - \bar D^0$ and $D^{*+}-D^{*-}$ yields.
This fact speaks well for the essential role
 of valence quarks of the initial hadron 
 even in the  region of  photon fragmentation at their photon energy.

The observed asymmetry cannot be explained
in the frame work of pQCD and factorization theorem for the 
inclusive  $D$-meson spectra. It is well known, 
that in the leading order on $\alpha_s$ 
of perturbative
theory  $D$- and $\bar D$-meson 
one gets identical spectra. At
 NLO calculations~\cite{NLO} it is appeared a small asymmetry 
 due to the quark-photon interaction:
\begin{equation}
\gamma + q\to c +\bar c+q.
\end{equation}

Indeed, the NLO-calculations for the photon energy 
of $E_\gamma=200$~GeV leads to the 
  asymmetry value which is smaller than the experimental one by the 
order of magnitude.  Thus, the hard production of $c\bar c$-pair followed by
the fragmentation into hadrons can not describe a such asymmetry.
It is worth to note that this situation is analogous to the situation with
weak decays of $D$-meson, where spectator mechanism can not explain lifetime 
difference between charged and neutral $D$-meson. So, one need to take 
into account the interaction with a charmless component of meson, as well
the interference terms.

Besides of an experimentally observed asymmetry there is  
another problem in using the factorization approach to
estimate $D$-meson production cross section.      
The thing is that at small transverse momenta one can not
  represent $D$-meson
production as $c\bar c$-pair production with further $c$-quark 
fragmentation into
$D$-meson, because  there is an interference between the light quark produced 
during $c$-quark fragmentation and that from a hadronic remnant. 
That taking into account  the interference contribution one could 
 determine the  region, where the factorization approach is valid.

We have tried to explain in  \cite{BKL} the both discussed effects 
(the asymmetry and the interference) 
 by including into  perturbative calculation in addition to heavy quarks the
 light ones needed to obtain a hadron with desired quantum numbers. 
In so doing the light quark has a mass $m_q\sim
\Lambda_{\mbox{\small{QCD}}}\sim 300$~MeV, which plays a role of 
an infrared cut-off. However, the inclusion of  
 the light one essentially complicates calculations. For example,  
the production process in  gluon-gluon interaction is described by 36 diagrams  
of $O(\alpha^4_s)$ order, 
 the production process in photon-photon interaction is 
 described by 24 diagrams
of  $O(\alpha\alpha_s^3)$ order.

The principal features of the discussed model are as follows:
\begin{enumerate}
\item
At large transverse momenta the model predictions coincide with the predictions 
of the fragmentation mechanism.  The cross section dependence on $D$-meson
transverse momentum $P^D_{\perp}$  have the standard factorized form:  
\begin{equation}
\frac{d\sigma (ij\to D +X)}{dp^D_{\perp}}=
\int \frac{d\sigma (ij\to c\bar c)}{dp^c_{\perp}}
f_{c\to D}(z_D)
\delta \left (z_D-\frac{p^D_{\perp}}{p^c_{\perp}}\right )
dz_Ddp^c_{\perp}
\label{factorization}
\end{equation}
were $i$ and $j$ are interacting partons,
 $p^c_{\perp}$ is a transverse momentum of $c$-quark, and
$d\sigma (ij\to c\bar c)/dp_c^{\perp}$
is the production cross section of $c\bar c$-pair in  
the leading order.
\item
In small transverse momenta region there is appeared
a new contribution that leads to power corrections  $\displaystyle
\frac{m^2_c}{p^6_T}$.
This contribution violates the factorized equation (\ref{factorization}).
\item
At small $p_T$ and large longitudinal momenta (i.e. at large $x$) it is
appeared a strong interference and the production cross section can not be
described by (\ref{factorization}).
\end{enumerate}

Note that the fragmentation function $f_{c\to D} (z_D)$ in
(\ref{factorization}) is calculated perturbatively  in the frame work of the 
discussed model and has the form: 
\begin{equation}
{f}_{c\to D^{(*)}} = 
\frac{\alpha^2_s \langle O (^3S_1)\rangle}{m_q^2m_c}
I^{D^{(*)}}(z_D,r),
\label{fact_view}
\end{equation}
where   $r=m_q/m_c$ and  $I$ is a function on $z_D$ and $r$. 
Let us to note, that the evaluated function   is close in form to
 Peterson fragmentation function~\cite{Peterson}.
\footnote{ It is important to stress, that the common  belief in
nonperturbative nature of Peterson fragmentation function is not valid. 
The dependence on $z$ in that function is caused by the heavy quark propagator. 
It had been shown in the original work~\cite{Peterson}, that to obtain this  
dependence one need to take into account only the right pole of 
propagator $1/(E-E')$ in the infinite reference frame.}

Variables  $\langle O(^3S_1)\rangle$ and $ \langle(O^1S_0)\rangle$ 
in the relativistic limit correspond to a wave function squared
at the coordinate origin. In the discussed model these variables
are considered as free parameters. The values of 
$\langle O(^3S_1)\rangle$ and $ \langle(O^1S_0)\rangle$ have been chosen 
to satisfy the normalization condition
\begin{equation}
W_{c\to D^{(*)}}=\int f_{c\to D^{(*)}}\,dz_D,
\label{normalization}
\end{equation}
where $W_{c\to D^{(*)}}$ is the probability of fragmentation 
$c \to D^{(*)}$ taken from the experimental date on $e^+e^-$-annihilation. 
As  it is seen from  (\ref{fact_view}) and (\ref{normalization}), the
values of $m_q$, $\alpha_s$ and $\langle O(^3S_1) \rangle  $ (or 
$\langle O(^1S_0) \rangle$) are correlated to each other. 
For example for $m_q=0.3$~GeV and  $\alpha_s=0.3$~GeV one gets
$$\langle O(^3S_1)\rangle 
\simeq \langle O(^1S_1) \rangle  = 0.25\,\mbox{GeV}^3.$$
These values allow to describe  the
experimental data on photo- and electroproduction  
at HERA collider~\cite{ZEUS}. 

It is worth to emphasize the role of the following process
 for the charm production:
\begin{equation}
\gamma +q\to D^{(*)} (\bar cq)+c.
\label{pr_process}
\end{equation}
Due to $q\leftrightarrow \bar q$ asymmetry in the structure functions of the
initial hadrons the process (\ref{pr_process}) leads to a charge asymmetry
 in   the $D^{(*)}$-meson yields. 
 This process dominates at low energies and large values of
Feynman variable   $x_F=\frac{2p_L^D}{\sqrt s}$ where 
$p_L^D$ is a longitudinal momentum of $D$-meson. 
At high energies at small $x_F$ the contribution of such process
 is negligible.  At kinematic conditions 
of HERA collider the process (\ref{pr_process}) contributes only 
several percents into the total cross section value, 
which is within experimental errors~\cite{BKL}. 

Following~\cite{Braaten} we refer  the process  (\ref{pr_process}) 
as a perturbative recombination.
The detailed analysis of the perturbative recombination reveals that
the cross section behavior  depends rather weakly on a light quark mass.
The analytic calculations  in the limit as $x_q\to 0$, 
where $x_q$ is a momentum fraction of $D$-meson carried by the light quark,
exhibit several interesting features of the process  (\ref{pr_process}).
That, accordingly to the results of~\cite{Braaten}, in the initial light quark
direction $(\theta=0)$
\begin{equation}
\left.\frac{d\sigma (\gamma+q\to (\bar cq)+c)}
{d\sigma (\gamma+q\to c+\bar c)}
\right|_{\theta=0}
\simeq
\frac{256 \pi}{81}\alpha_s
\label{quark_direction}
\end{equation}
for both $(\bar cq)(^1s_0)$ and  $(\bar cq)(^3s_1)$-states. One can see
that in spite of an additional power of $\alpha_s$ the cross section value 
of the process (\ref{pr_process}) is comparable to the cross section value
of the photon-gluon fusion.
In the photon direction $(\Theta=\pi)$  the production cross section
is  suppressed
by additional factors of $m^2_c/s$~\cite{Braaten}:
\begin{equation}
\left.\frac{d\sigma (\gamma+q\to (\bar cq)+c)}
{d\sigma (\gamma+g\to \bar c+ c)}
\right|_{\theta=\pi}
\simeq
\begin{array}{lcr}
\frac{256 \pi}{81}\alpha_s\frac{m^6_c}{s^3} & \mbox{for} & ^1S_0
\\  [0.4cm]
\frac{256 \pi}{81}\alpha_s\frac{m^2_c}{s} & \mbox{for} & ^3S_1
\end{array}.
\label{photon_direction}
\end{equation}

At large transverse momenta $p_\perp$ the cross section is suppressed by 
an additional factor of   $m_c^2/p^2_\perp$ and falls of like 
 $\displaystyle\frac{1}{p^6_\perp}$.

\section{Calculation results}

\begin{figure}
{\centering 
\resizebox*{\textwidth}{!}
{\includegraphics{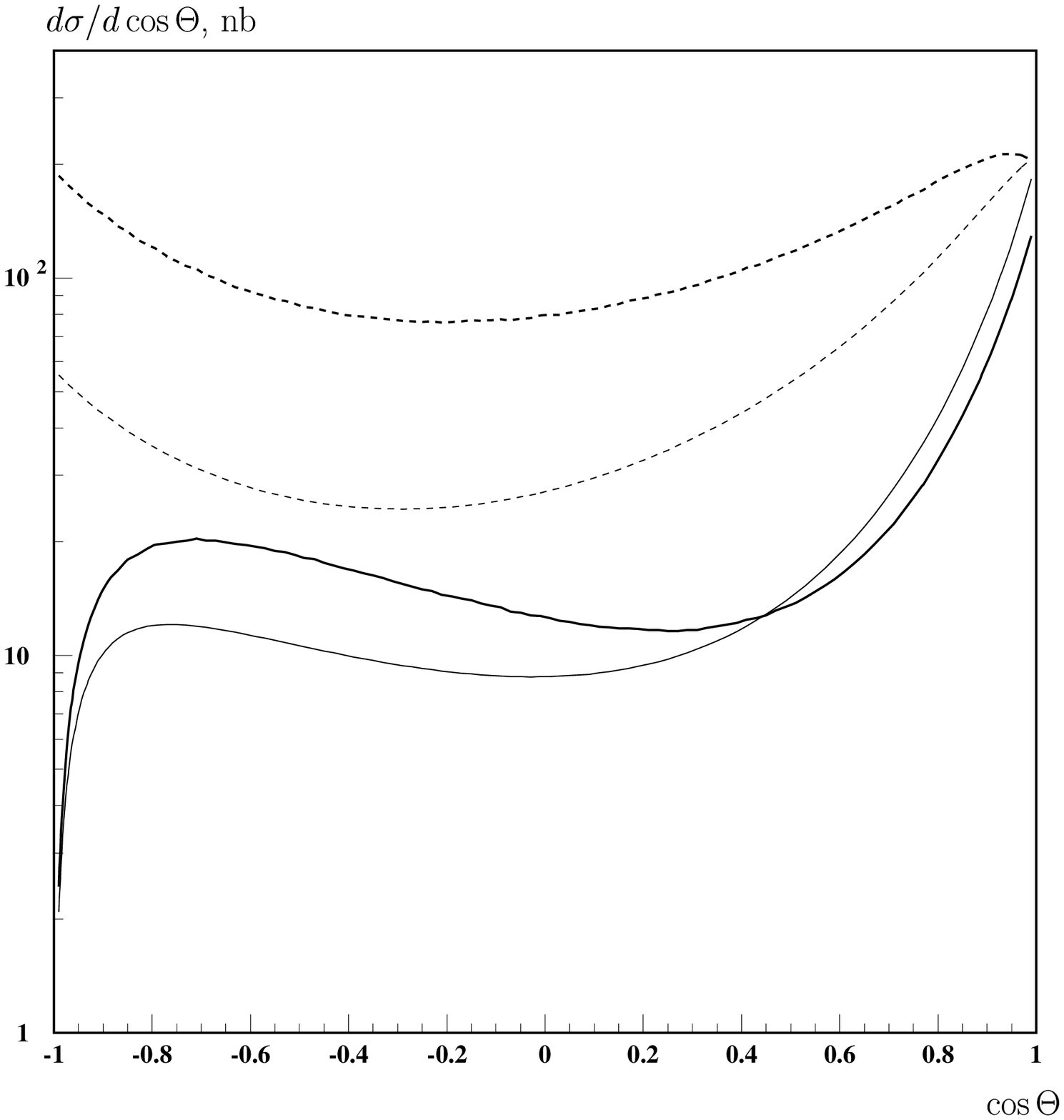}} 
\vspace*{-5cm} \par}
\setcaptionmargin{5mm}
\onelinecaptionsfalse
\captionstyle{normal}
\caption{
 The cross section distributions over the 
scattering angle cosine of \((d\bar c)\)-meson in c.m.s. for the process
 \(d\gamma\to (d\bar c) +c\) both for the  vector (solid curves) and
pseudoscalar (dashed curves) meson states.
Ours predictions (bold curves) have been performed 
in comparison with  the predictions of work~\cite{Braaten} (thin curves).    
\(\sqrt{s_{d\gamma}}=5\)~GeV, \(m_c=1.5\)~GeV, \(m_d=0.3\)~GeV.
}
\label{theta_real}
\end{figure}

\begin{figure} 
{\centering 
\resizebox*{\textwidth}{!}
{\includegraphics{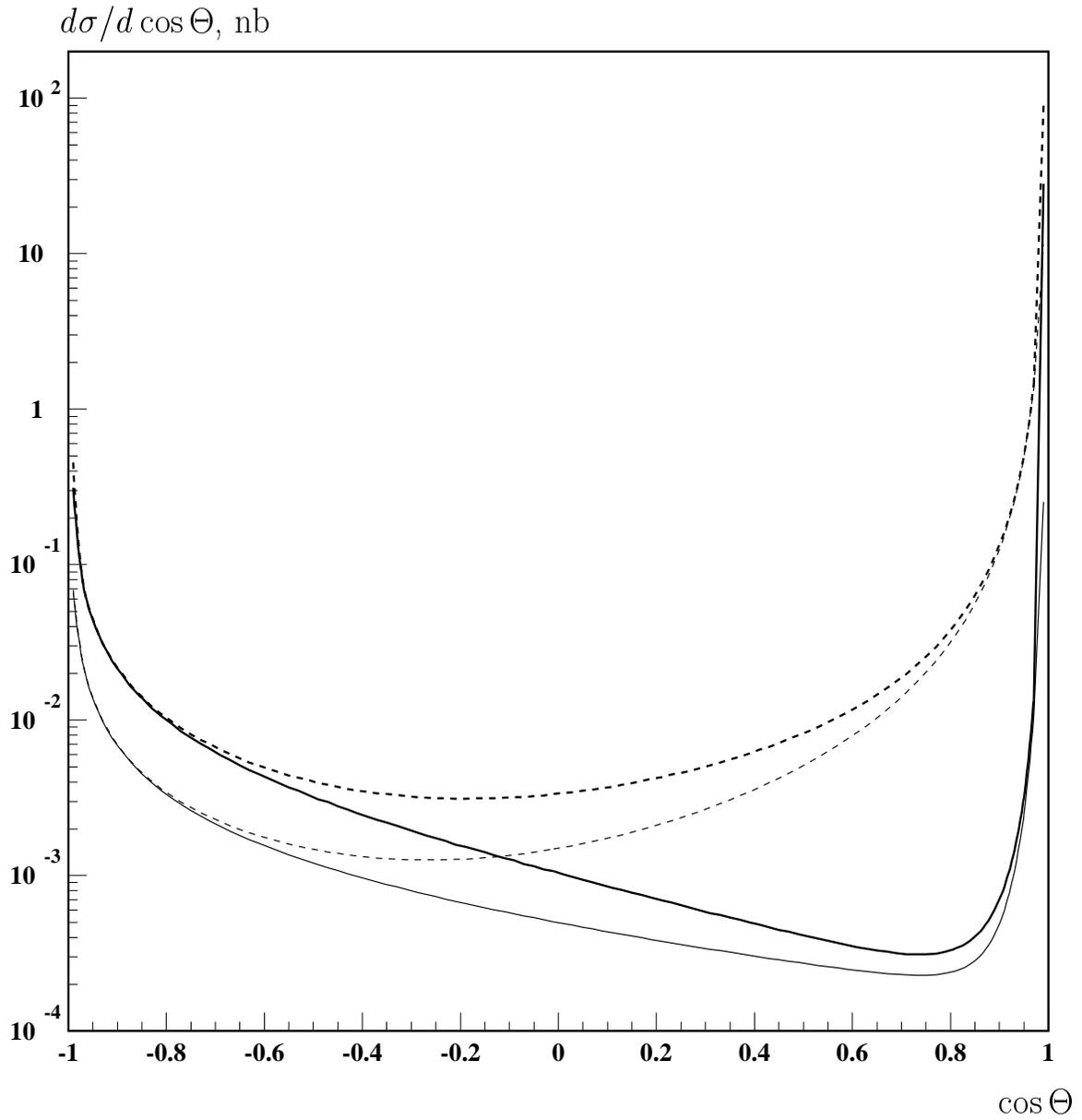}} 
\vspace*{-5cm} \par}
\setcaptionmargin{5mm}
\onelinecaptionsfalse
\captionstyle{normal}
\caption{
The same as in Fig.~\ref{theta_real} but for 
\(\sqrt{s_{d\gamma}}=100\)~GeV, \(m_c=1.5\)~GeV, and \(m_d=0.1\)~GeV.
}
\label{theta_limit}
\end{figure}

The cross section dependence on scattering angle for perturbative
recombination is shown in 
 Fig.~\ref{theta_real}for both   vector and pseudoscalar 
states of  $(\bar cq)$-system.
For the sake of comparison   
 the cross section behaviour is presented for the case of 
limit  $x_q\to 0$~\cite{Braaten}, as well as 
for the nonzero mass of light quark $m_q=0.3$~GeV. 
One can see that at low energies the 
 production cross section for pseudoscalar $(\bar cq)$-state 
in the photon direction $(\Theta=\pi)$  is more then 
10 times  less than the production cross section 
of vector $(\bar cq)$-state.
These calculation results become similar to each other at high energy and 
in the limit as $m_q \to 0$ (see Fig.~\ref{theta_limit}).
It is seen from these figures that  the $(\bar cq)$-state production   
 in the backward  direction (in the photon direction) is suppressed 
in comparison with the production in the forward direction 
(the factor of $m^2(s)$).
 As a rule at fixed target facilities  the research of production
is only possible in the photon direction. 
 Therefore the contribution of perturbative recombination must be small.  
 In addition to that the charge asymmetry should decrease
 with the increasing of
beam  energy. On the other hand there is no suppression 
at the production yield alongside to hadron direction
 (see eq.~(\ref{quark_direction})). 
It leads to energy-independent charge asymmetry.  

Before proceeding further, 
it is necessary to remove perplexity
which could appear after comparing  the predictions 
(\ref{quark_direction}) and (\ref{photon_direction}) with the results
following from Fig.~\ref{theta_limit}.
Indeed one can see from Fig.~\ref{theta_limit}  that  the yields 
of vector and  pseudoscalar mesons
 are practically identical in the backward production 
($\Theta\simeq \pi$), while  
 the vector mesons dominate in forward production ($\Theta\simeq 0$). 
However, these results contradict   
(\ref{quark_direction}) and (\ref{photon_direction}).
The point is that the  mathematical limits of the cross section values evaluated
in~\cite{Braaten} are correct. But these quantities have no physical meaning.
As it is shown in our study the production cross sections
for vector and pseudoscalar states are practically the same 
in the  kinematical region $-0.99<\cos \Theta<-0.8$, whereas the limit regime
establishes only at  $\cos \Theta<-0.99$.
The integral cross section value in the kinematical region
 $-1< \cos \Theta<-0.99$
is negligible  in comparison with the integral cross section 
in the kinematic region $-0.99<\cos \Theta<-0.8$.
It was revealed also that in the forward production at  
$0.9<\cos \Theta<0.999$  the vector mesons dominate: while
at  $\cos \Theta =1$ the ratio between the vector meson production and 
the pseudoscalar production  equals to 1, but at  $\cos \Theta =0.999$
that ratio is about 7.
The contributions from region  $0.999<\cos \Theta<1$,
as well as from region $-1< \cos \Theta<-0.99$ into the total cross
section are negligible.
Now it is clear that the predictions 
(\ref{quark_direction}) and  (\ref{photon_direction}) are beyond both
the model accuracy and the experimental  potentialities. 
Therefore the  physically reasonable predictions at large $s$
are as follows:
\begin{equation}
\left.\frac{d\sigma (\gamma+q\to (\bar cq)(^3s_1)+c)}
{d\sigma (\gamma+q\to (\bar cq)(^1s_0)+c)}
\right|_{\theta\simeq 0, \, s\gg m_c^2} \gg 1,\;
\left.d\sigma (\gamma+q\to (\bar cq)+c)
\right|_{\theta\simeq 0, \, s\gg m_c^2}\sim \frac{1}{s},
\end{equation}

\begin{equation}
\left.\frac{d\sigma (\gamma+q\to (\bar cq)(^3s_1)+c)}
{d\sigma (\gamma+q\to (\bar cq)(^1s_0)+c)}
\right|_{\theta\simeq \pi, \, s\gg m_c^2}\simeq 1, \;
\left.d\sigma (\gamma+q\to (\bar cq)+c)
\right|_{\theta\simeq \pi, \, s\gg m_c^2}\sim \frac{1}{s^4}.
\end{equation}

Let us consider  in detail the both perturbative recombination and 
 photon-gluon fusion.         
We restrict ourself  to the calculation of production
asymmetry for  $D^*$-mesons, because we expect the suppression of the 
 $D$-meson production in the photon direction 
in comparison with the $D^*$-meson production.  

For the numerical analysis we choose the kinematical cuts of
COMPASS experiment, t.e.: the averaged energy of the muon beam is 
160 GeV, the photon virtuality  $Q^2< 1$~GeV$^2$;
$x_F>0.2$, where $x_F$ is 
the Feynman variable for  $D^*$-meson;
$z_{\gamma}>0.2$,  where   $z_{\gamma}$ is the fraction of 
photon momentum carried out by  $D^*$-meson.

\begin{figure} 
{\centering 
\resizebox*{\textwidth}{!}
{\includegraphics{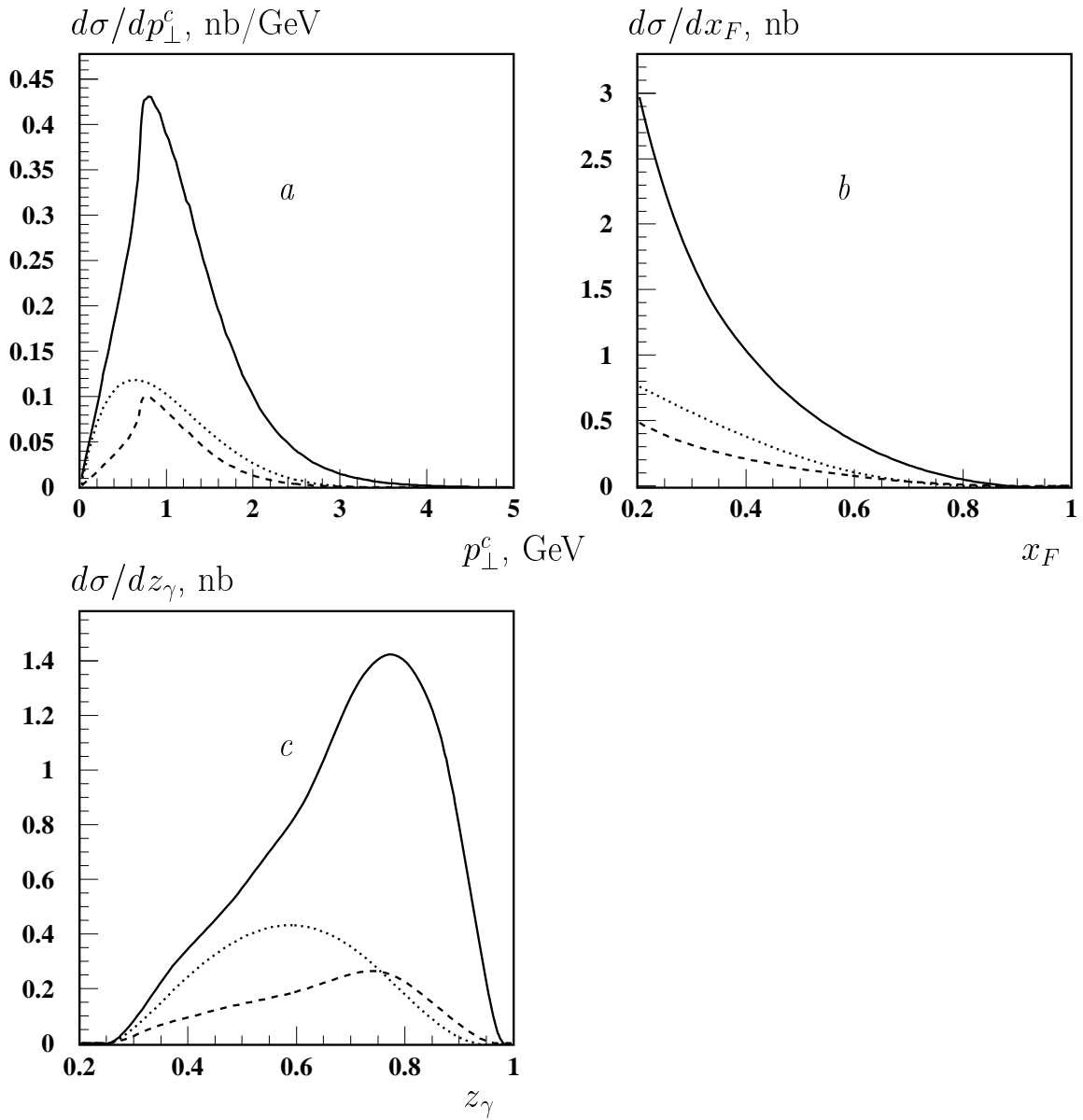}} 
\vspace*{-4cm} \par}
\setcaptionmargin{5mm}
\onelinecaptionsfalse
\captionstyle{normal}
\caption{
The cross sections as a function of
 \(p^D_{\perp}\)~(a), 
\(x_F\)~(b), and \(z_{\gamma}\)~(c) for the following
 subprocesses of charged charmed meson photoproduction: 
 \(D^{*-}\)-meson production in the
perturbative recombination process \(d \gamma \to D^{*-}+ c\) (solid
curve); \(D^{*+}\)-meson production in the
perturbative recombination process \(\bar d \gamma \to D^{*+}+ \bar c\) (dashed
curve);
the production of \(D^{*+}\) and \(D^{*-}\)-mesons in photon-gluon fusion.
}
\label{cd_recombination}
\end{figure}

\begin{figure} 
{\centering 
\resizebox*{\textwidth}{!}{\includegraphics{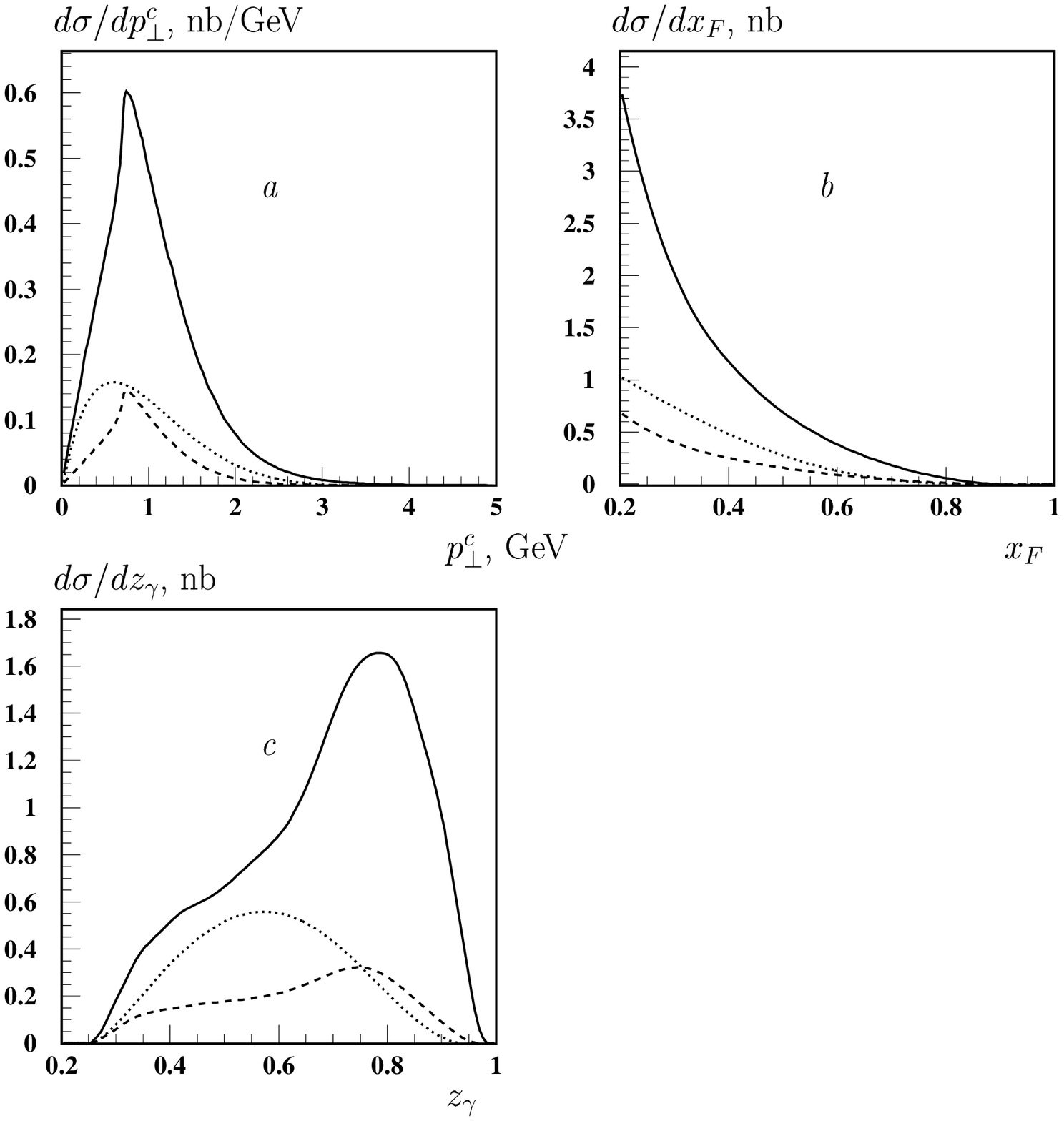}}
\vspace*{-5cm} \par}
\setcaptionmargin{5mm}
\onelinecaptionsfalse
\captionstyle{normal}
\caption{
The cross sections as a function of
 \(p^D_{\perp}\)~(a), 
\(x_F\)~(b), and \(z_{\gamma}\)~(c) for the following
 subprocesses of neutral charmed meson photoproduction: 
\(\bar D^{*0}\)-meson production in the
perturbative recombination process \(u \gamma \to \bar D^{*0}+ c\) (solid
curve);
\(D^{*0}\)-meson production in the
perturbative recombination process \(\bar u \gamma \to D^{*0}+ \bar c\) 
(dashed curve);
the production of \(D^{*0}\) and \(\bar D^{*0}\)-mesons in photon-gluon fusion.
}
\label{cu_recombination}
\end{figure}

Fig.~\ref{cd_recombination} presents the  cross sections for
$D^{*+}$ and $D^{*-}$-meson production
calculated in the frame work of the  perturbative recombination mechanism
 has been shown as well as 
the contribution due to  photon-gluon fusion. 
The perturbative recombination
contributes to the charge asymmetric part of the total cross section,
while the  photon-gluon fusion contributes to the charge symmetric one.
The predictions of perturbative
recombination and photon-gluon fusion  for   
 $D^{*0}$ and  $\bar D^{*0}$-meson production are presented in  
 Fig.~\ref{cu_recombination}.

\begin{figure} 
{\centering \resizebox*{\textwidth}{!}{\includegraphics{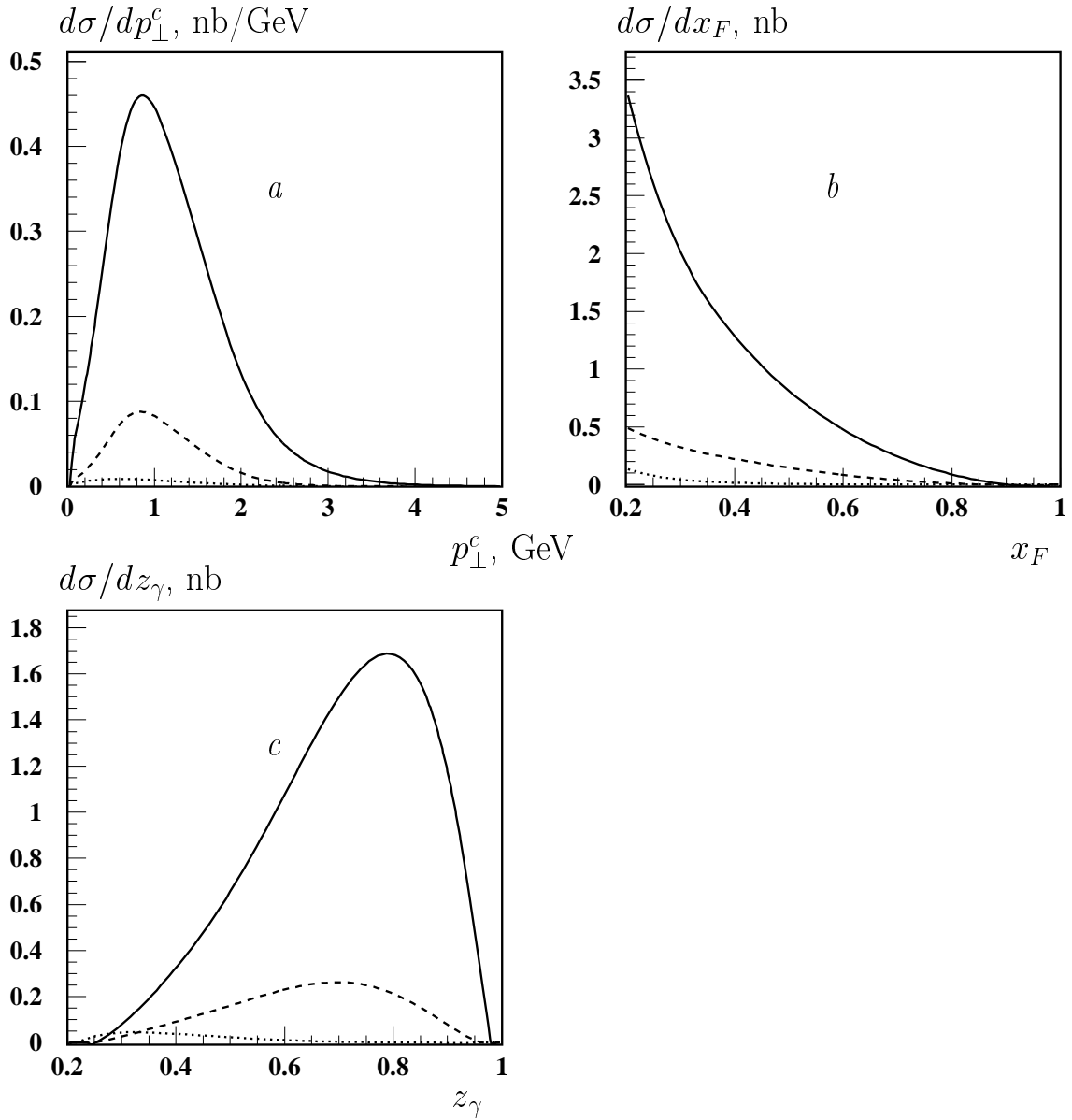}}
\vspace*{-5cm} \par}
\setcaptionmargin{5mm}
\onelinecaptionsfalse
\captionstyle{normal}
\caption{
The cross sections as a function of the kinematic variables of
"nonrecombinated" charm quark 
 \(p^c_{\perp}\)~(a), 
\(x^c_F\)~(b), and \(z^c_{\gamma}\)~(c) for the following
 subprocesses of charged charmed meson photoproduction: 
 \(D^{*-}\)-meson production in the
perturbative recombination process \(d \gamma \to D^{*-}+ c\) (solid
curve); \(D^{*+}\)-meson production in the
perturbative recombination process \(\bar d \gamma \to D^{*+}+ \bar c\)(dashed
curve);
the production of \(D^{*+}\) and \(D^{*-}\)-mesons in photon-gluon fusion.
}
\label{cd_recombination_c_quark}
\end{figure}

\begin{figure} 
{\centering \resizebox*{\textwidth}{!}{\includegraphics{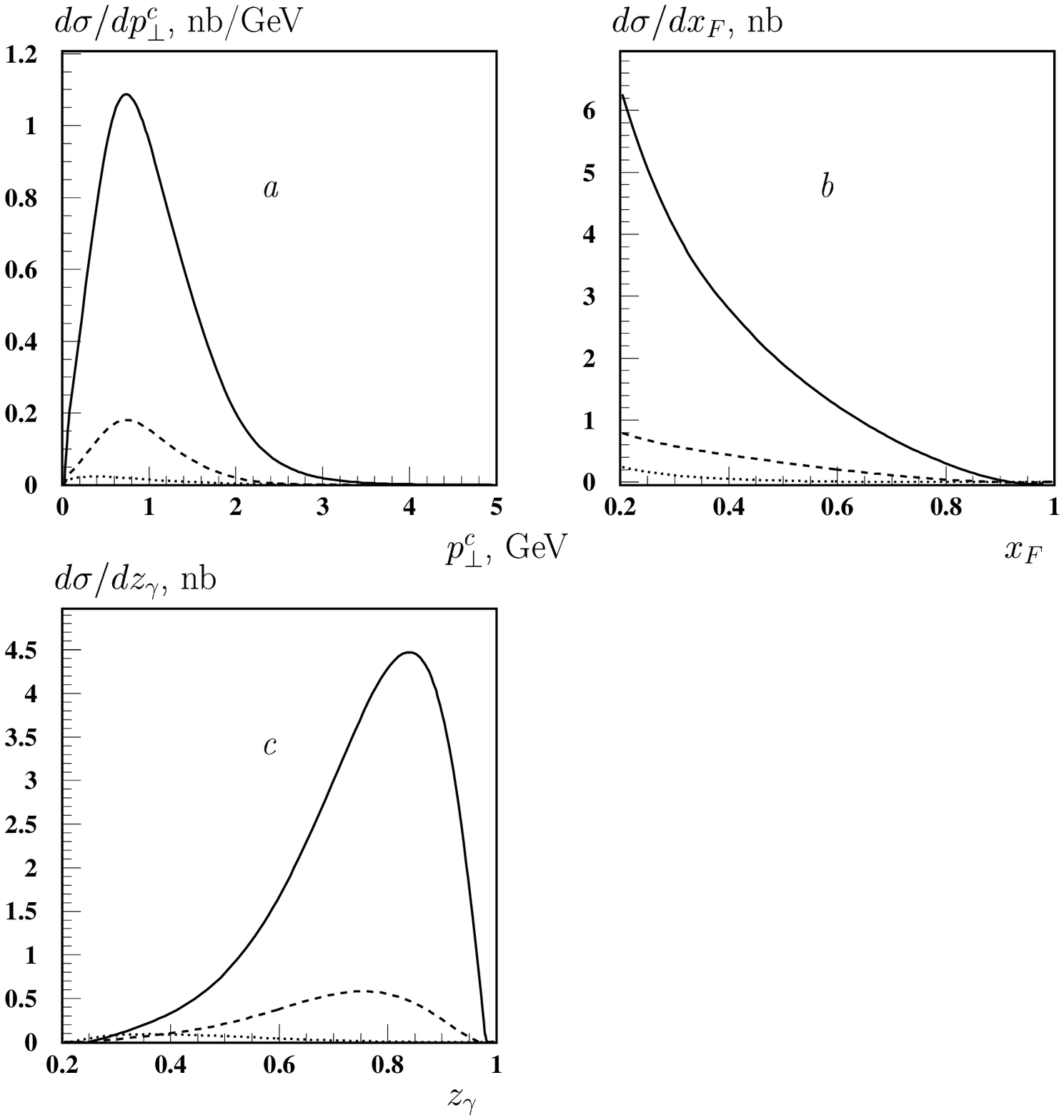}}
\vspace*{-5cm} \par}
\setcaptionmargin{5mm}
\onelinecaptionsfalse
\captionstyle{normal}
\caption{
The cross sections as a function of the kinematic variables of
"nonrecombinated" charm quark  \(p^c_{\perp}\)~(a), 
\(x^c_F\)~(b), and \(z^c_{\gamma}\)~(c) for the following
 subprocesses of neutral charmed meson photoproduction:
\(\bar D^{*0}\)-meson production in the
perturbative recombination process \(u \gamma \to \bar D^{*0}+ c\) (solid
curve);
\(D^{*0}\)-meson production in the
perturbative recombination process \(\bar u \gamma \to D^{*0}+ \bar c\) 
(dashed curve);
the production of \(D^{*0}\) and 
\(\bar D^{*0}\)-mesons in photon-gluon fusion.
}
\label{cu_recombination_c_quark}
\end{figure}

In the kinematical region under consideration the perturbative recombination
dominates due to a valence quark of the initial proton, as it is seen
 in Fig.~\ref{cd_recombination} and \ref{cu_recombination}.
On the face of it, it seems that   $\sigma(D^-)\ll\sigma(D^+)$ and 
$\sigma(D^0)\ll\sigma(\bar D^0)$. However, it is worth to note that
 $D^-$($\bar D^0$)-meson  in the perturbative recombination 
is produced in association with the $c$-quark
(see Fig.~\ref{cd_recombination_c_quark} and \ref{cu_recombination_c_quark}).
To estimate the charge asymmetry value resulted 
from from a pertubative recombination one needs to keep in mind that
$c$-quark also produces the charm particles.
Let us suppose, that $c$-quark hadronizes into final states with probabilities
$W_{D^*}$ ($c \to D^*$),   $W_{D}$ ($c \to D$), $W_{D_s}$ ($c \to D_s$) and 
$W_{\Lambda_c}$ ($c \to \Lambda_c$) (each do not necessarily equal to
the appropriate probabilities at large $c$-quark momenta). To calculate the 
contribution of $c$-quark factorization process into the cross section one 
needs to take into account only particles, which are produced by $c$-quark
 in the kinematical region of COMPASS experiment.    

To explain the assertion stated above, let us consider an example. 
To estimate  the yield of the $D^{*-}$-meson produced in the
recombination  process, $\gamma d \to D^{*-}(\bar c d)+c$,  we
apply the following cuts $x_F(D^{*-})>0.2$, $z_{\gamma}(D^{*-})>0.2$ 
and do not
care about $c$-quark. To estimate the contribution of the remained 
$c$-quark, then we apply the cuts  on $x_F>0.2$ and  $z_{\gamma}>0.2$ for the
product of $c$-quark hardronization and do not care about 
$D^{*-}(\bar c d)$-meson. 

Since an interaction energy is closed to threshould one,
it is reasonable to suppose that the total momentum of $c$-quark carried out
by the charm meson.
 However, one can not exclude that the charm meson carries out only a part of the 
$c$-quark momentum, as it takes place at high interaction energies.
That is why we consider two  extreme cases: 
\begin{enumerate}
\item the total momentum of  $c$-quark is passed to meson; 
\item An energy loss of $c$-quark is described by
Kartvelishvili fragmentation function \cite{Kartvelishvili}:
\begin{equation}
f(z) \sim z_D^{2.2}(1-z_D)
\end{equation}
\end{enumerate} 

The charge asymmetry of $D^{*-}$-meson production is described by the equation:

\begin{equation}
A^{D^{*+}}=\frac{\sigma^{D^{*+}}_{\rm total}-\sigma^{D^{*-}}_{\rm total}}
{\sigma^{D^{*+}}_{\rm total}+\sigma^{D^{*-}}_{\rm total}}.
\end{equation}

Let us $\sigma^D_{\gamma g}$ be the cross section value for meson production
 in the photon-gluon fusion;  $\sigma^D_{\gamma q}$ is
the cross section value of meson production in the 
perturbative recombination with light quark $q$;
while $\sigma^c_{\gamma q}$ is the cross section value of
 the  "nonrecombinated" $c$-quark   production.
 Using these notations the asymmetry has the form as follows:
\begin{equation}
A^{D^{*+}}=\frac{(
\sigma^{D^{*+}}_{\gamma \bar d}-\sigma^{D^{*-}}_{\gamma d})
-W_{D^*}(
\sigma^{\bar c}_{\gamma \bar d}-\sigma^{c}_{\gamma d}+
\sigma^{\bar c}_{\gamma \bar u}-\sigma^{c}_{\gamma u})
}
{
(
\sigma^{D^{*+}}_{\gamma \bar d}+\sigma^{D^{*-}}_{\gamma  d})
+W_{D^*}(
\sigma^{\bar c}_{\gamma  \bar d}+\sigma^{c}_{\gamma d}+
\sigma^{\bar c}_{\gamma \bar u}+\sigma^{c}_{\gamma u})+
2\sigma^{D^{*+}}_{\gamma g}}.  
\end{equation}

For the case (1) when the meson carries out the whole $c$-quark momentum
our calculation leads to the following result: 
$$ A^{D^{*+}}_{1} \simeq -0.03. $$

For the case (2) when an energy loss of $c$-quark is described by
Kartvelishvili fragmentation function \cite{Kartvelishvili}, one gets: 
$$ A^{D^{*+}}_{2} \simeq -0.17. $$

For the neutral meson production one has another predictions:
$$ A^{D^{*0}}_{1}\simeq -0.08, $$
$$ A^{D^{*0}}_{2}\simeq -0.21. $$

One can see that the asymmetry values both for neutral and charged mesons 
strongly depend on assumptions  about the fragmentation mechanism for the
"nonrecombinated" $c$-quark. Therefore  a lack of knowledge about 
this mechanism does not allow to calculate asymmetry values 
 $A^{D^{*+}}$ and $ A^{D^{*0}}$ with reasonable
accuracy. 

Nevertheless, the more accurate prediction really exits in the
considered kinematical
region. In spite of the absence 
of  $s$-quark inside an initial nucleon the perturbative mechanism leads to
the charged asymmetry in  $D_s$-meson production.
As it  mention above, the production of the mesons with 
$\bar c$-quark in the perturbative recombination occurs
with an additional  production of 
$c$-quark. This $c$-quark can hadronize into  $D_s^+$.  Therefore
an additional yield of $D^{*-}$ and  $\bar D^{*0}$-mesons leads
to an additional yield of  $D_s^+$-mesons. 

The asymmetry for $D_s$-meson production is described by following expression:
\begin{equation}
A^{D_s^+}=\frac{-W_{D_s}
(
\sigma^{\bar c}_{\gamma \bar d}-
\sigma^{c}_{\gamma  d}
+\sigma^{\bar c}_{\gamma \bar u}
-\sigma^{c}_{\gamma u}
)}
{W_{D_s}(
 \sigma^{\bar c}_{\gamma \bar d}
+\sigma^{c}_{\gamma d}
+\sigma^{\bar c}_{\gamma \bar u}
+\sigma^{c}_{\gamma u})+
2\sigma^{D_s^+}_{\gamma g}}.  
\end{equation}

Our estimates are given below:
$$A^{D_s^+}_{1}\simeq 0.57$$ and
$$A^{D_s^+}_{2}\simeq 0.64$$

Thus, the perturbative recombination mechanism leads to the essential charge 
asymmetry in the process of  $D_s$-meson production. The value of such 
asymmetry depends weakly
  on  assumptions  about the fragmentation mechanism for the
"nonrecombinated" $c$-quark.

\section{Conclusion}
The perturbative recombination mechanism, 
which violates a factorization theorem in charmed 
hadron production causes the charge asymmetry in 
production of these particles.
To calculate correctly the asymmetry value, ones need to take into
account the hardonization of  "nonrecombinated" charm (anticharm) quark, 
which is produced    together  with the  anticharm (charm) quark to
be recombinated with the light constituent quark of initial proton.

In the considered process of the charm photoproduction into
the region of photon fragmentation the nonzero value of charge asymmetry
takes place due to the backward scattering of valence quark of the initial
hadron. In the discussed kinematical region the asymmetry becomes negligible 
with energy increasing. 
In the condition of COMPAS experiment the asymmetry value is small
in $D^\star$-meson production, but it is essential in $D_s$-meson production.

Therefore one can observe an interesting phenomenon:
the production of $D^{+(\star)}$ meson,  which contains 
 the  light valence quark from the  initial
proton,  leads 
to the charge asymmetry in the  production of $D_s$-mesons, which does not 
contain the valence quark from the initial proton.     
In spite of small $D_s$ meson yield, the  charge asymmetry $ A^{D_s^+}$
is large and it depends insignificantly on details of a fragmentation model.
Note that the first time a such effect was considered in~\cite{Zaitsev}
where a possible asymmetry  in $B_s$-meson production in $pp$-collisions
was investigated. 
This process  had been  also investigated  in paper~\cite{Braaten}.

The authors are grateful to  S.~V.~Donskov and  S.~R.~Slabospitsky 
for the fruitful discussion and the help.

This work is partially supported 
by Grants of RFBR (04-02-17530),  
CRDF (M0-011-0) and has been  realized in the frame work of
scientific school  1303.2003.2.

\newpage

\end{document}